\documentclass[a4paper,10pt]{sigwebnewsletter}

\let\pdfbookmark\relax

\usepackage{xcolor}
\usepackage{amsfonts}
\usepackage{amsthm}
\usepackage{booktabs}
\usepackage{graphicx}
\usepackage{helvet}
\usepackage{courier}
\usepackage{makecell}
\usepackage{float}
\usepackage{bm}
\usepackage{multirow}
\usepackage{hyperref}
\usepackage{flushend}
\usepackage{bbm}
\usepackage{mathrsfs}
\usepackage{enumitem}
\theoremstyle{definition}
\newtheorem{definition}{Definition}[section]

\title{Deep Reinforcement Learning for Search, Recommendation, and Online Advertising: A Survey}

\author{Xiangyu Zhao, Michigan State University \\ Long Xia, JD.com \\ Jiliang Tang, Michigan State University \\ Dawei Yin, JD.com}

\newsletterQuarter{Spring}
\newsletterYear{2019}

\begin{abstract}
Search, recommendation, and online advertising are the three most important information-providing mechanisms on the web. These information seeking techniques, satisfying users' information needs by suggesting users personalized objects (information or services) at the appropriate time and place, play a crucial role in mitigating the information overload problem. With recent great advances in deep reinforcement learning (DRL), there have been increasing interests in developing DRL based information seeking techniques. These DRL based techniques have two key advantages -- (1) they are able to continuously update information seeking strategies according to users' real-time feedback, and (2) they can maximize the expected cumulative long-term reward from users where reward has different definitions according to information seeking applications such as click-through rate, revenue, user satisfaction and engagement. In this paper, we give an overview of deep reinforcement learning for search, recommendation, and online advertising from methodologies to applications, review representative algorithms, and discuss some appealing research directions.

\end{abstract}
\begin{document}
\maketitle
\section{Introduction}
The explosive growth of the World Wide Web has generated massive data. As a consequence, the information overload problem has become progressively severe~\cite{DBLP:journals/tkde/ChangKGS06}. Thus, how to identify objects that satisfy users' information needs at the appropriate time and place has become increasingly important, which has motivated three representative information seeking mechanisms -- search, recommendation, and online advertising.  The search mechanism outputs objects that match the query, the recommendation mechanism generates a set of items that match users' implicit preferences, and the online advertising mechanism is analogous to search and recommendation expect that the objects to be presented are advertisements~\cite{DBLP:journals/cacm/Garcia-MolinaKP11}. Numerous efforts have been made on designing intelligent methods for these three information seeking mechanisms. However, traditional techniques often face several common challenges. First, the majority of existing methods consider information seeking as a static task and generate objects following a fixed greedy strategy. This may fail to capture the dynamic nature of users’ preferences (or environment). Second, most traditional methods are developed to maximize the short-term reward, while completely neglecting whether the suggested objects will contribute more in long-term reward~\cite{DBLP:journals/jmlr/ShaniHB05}. Note that the reward has different definitions among information seeking tasks, such as click-through rate~(CTR), revenue, and dwell time.

Recent years have witnessed the rapid development of reinforcement learning (RL) techniques and a wide range of RL based applications. Under the RL schema, we tackle complex problems by acquiring experiences through interactions with a dynamic environment. The result is an optimal policy that can provide solutions to complex tasks without any specific instructions~\cite{DBLP:journals/jair/KaelblingLM96}. Employing RL for information seeking can naturally resolve the aforementioned challenges. First, considering the information seeking tasks as sequential interactions between an RL agent (system) and users (environment), the agent can continuously update its strategies according to users' real-time feedback during the interactions, until the system converges to the optimal policy that generates objects best match users’ dynamic preferences. Second, the RL frameworks are designed to maximize the long-term cumulative reward from users. Therefore, the agent could identify objects with small immediate reward but making big contributions to the reward in the long run.

Given the advantages of reinforcement learning, there have been tremendous interests in developing RL based information seeking techniques. Thus, it is timely and necessary to provide an overview of information seeking techniques from a reinforcement learning perspective. In this survey, we present a comprehensive overview of state-of-the-art RL based information seeking techniques and discuss some future directions. The remaining of the survey is organized as follows. In Section 2, we introduce technical foundations of reinforcement learning based information seeking techniques. Then we review the three key information seeking tasks -- search, recommendation, and online advertising -- with representative algorithms from Sections 3 to 5. Finally, we conclude the work with several future research directions. 
\section{Technical Foundations}
Reinforcement learning is learning how to map situations to actions
\cite{sutton1998introduction}.
The two fundamental elements in RL are to formulate the situations~(mathematical models) and to learn the mapping~(policy learning).

\subsection{Problem Formulation}

In reinforcement learning, there two main settings for problem formulations: multi-armed bandits~(without state transition) and Markov decision processes~(with state transition).

\subsubsection{Multi-Armed Bandits}
The Multi-Armed Bandits~(MABs) problem is a simple model for the exploration/exploitation trade-off~\cite{DBLP:conf/mcpr/VaraiyaW83}.
Formally, a $K$-MAB can be defined as follows.
\theoremstyle{definition}
\begin{definition}
A $K$-MAB is a 3-tuple $\langle A,R,\pi \rangle$, where $A$ is the set of actions~(arms) and $|A|=K$, $r=R(a)$ is the reward distribution when performing action $a$, and policy $\pi$ describes probability distribution over the possible actions.
\end{definition}
An arm with the highest expected reward is called the \textit{best arm}~(denoted as $a_*$)~and its expected reward $r_{*}$ is the \textit{optimal reward}.
An algorithm for MAB, at each time step $t$, samples an arm $a_t$ and receives a reward $r_t$. When making its selection, the algorithm depends on the history~(i.e., actions and rewards) up to the time $t$-$1$. The contextual bandit model~(a.k.a. associative bandits or bandits with side information) is an extension of MAB that takes additional information into account~\cite{DBLP:journals/ml/AuerCF02,DBLP:journals/jmlr/LuPP10}.

\subsubsection{Markov Decision Process}
A Markov decision process~(MDP) is a classical formalization of sequential decision making, which is a mathematically idealized form of reinforcement learning problem~\cite{bellman2013dynamic}.
We define an MDP as follows.
\theoremstyle{definition}
\begin{definition}
A Markov Decision Process is a 5-tuple $\langle S,A,T,R,\pi \rangle$, where $S$ is a set of states, $A$ is a discrete set of actions, $T$ is the state transition function $s_{t+1}=T(s_t,a_t)$ which specifies a function mapping a state $s_t$ into a new state $s_{t+1}$ in response to the selected action $a_t$, $r=R(s,a)$ is the reward distribution when performing action $a$ in state $s$, and policy $\pi(a|s) $ describes the behaviors of an agent which is a probability distribution over the possible actions.
\end{definition}

The agent and environment interact at each of a sequence of discrete time steps $t=\{0,1,2,\dots\}$. Consequently, a sequence or \textit{trajectory} is generated as $\{s_0,a_0,r_1,\cdots,s_t,a_t,$\\$r_{t+1},\cdots\}$. In general, we seek to maximize the \textit{expected discounted return}, where the return $G_t$ is defined as: $G_t=\sum_{k=0}^{\infty}\gamma^{k}r_{r+k+1}$, where $\gamma$ ($0\leq\gamma\leq1$) is the \textit{discounted rate}.
\iffalse
The \textit{value function} of a state $s$ under a policy $\pi$, denoted $\pi_v(s)$, is formally defined as: $\pi_v(s)=\mathbb{E}_{\pi}[G_t|s]$, where $\mathbb{E}_{\pi}$ denotes the expected value of a random variable given that the agent follows policy $\pi$.
Similarly, the value of taking action $a$ in state $s$ under a policy $\pi$, denoted $q_{\pi}(s,a)$, as the expected return starting from $s$, taking the action $a$, and thereafter following policy $\pi$: $q_{\pi}(s,a)=\mathbb{E}_{\pi}[G_t|s,a]$.
$v_{*}(s)$ is the \textit{optimal state-value function}, and defined as the expected return on the \textit{optimal policy} $\pi_{*}$.
And the \textit{Bellman optimality equation} for $v_{*}(s)$ is defined as:
\[
v_{*}(s)=\max_{a}\sum_{s',r}p(s',r|s,a)[r+\gamma v_{*}(s')],
\]
where $p(s',r|r,a)$ is the \textit{state-transition probabilities}.
The Bellman optimality equation for $q_{*}$, the \textit{optimal action-value function}, is defined as:
\[
q_{*}(s,a)=\sum_{s',r}p(s',r|s,a)[r+\gamma \max_{a'}q_{*}(s',a')]
.\]
Intuitively, the Bellman optimality equation expresses the fact that the value of a state under an optimal policy must equal the expected return for the best action from that state, and vice versa.
\fi
%
The Partially Observable Markov Decision Process (POMDP) is an extension of MDP to the case where the state of the system is not necessarily observable~\cite{aastrom1965optimal,DBLP:journals/ior/SmallwoodS73,DBLP:journals/ior/Sondik78,DBLP:journals/ai/KaelblingLC98}.

\subsubsection{Multi-Agent Setting}
The generalization of the Markov Decision Process to the Multi-Agent case is the stochastic game~\cite{DBLP:journals/ai/BowlingV02,shoham2003multi,DBLP:journals/tsmc/BusoniuBS08} as: 
\theoremstyle{definition}
\begin{definition}
A multi-agent game is a tuple $\langle S,A_1,\dots,A_n,T,R_1,\dots,R_n,\pi_{1},\dots,\pi_{n}\rangle$, where $n$ is the number of agents, $S$ is the discrete set of environment states, $A_i$ is the discrete set of actions for the agent $i$, $T$ is the state transition probability function, $R_i$ is the reward function of agent $i$, and $\pi_{i}$ is the policy adopted by agent $i$. 
\end{definition}

In the multi-agent game, the state transition is the result of the joint actions of all the agents $\mathbf{a}_t=[a_{1,t}^T,\dots,a_{n,t}^T]^T$, where $a_{i,t}\in A_{i}$ denotes the action taken by agent $i$ at time step $t$.
The reward $r_{i,k+1}$ also depends on the joint action.
If $\pi_{1}=\cdots=\pi_{n}$, i.e., all the agents adopt the same policy to maximize the same expected return, the multi-agent game is fully cooperative.
If $n=2$ and $\pi_{1} = - \pi_{2}$, i.e., the two agents have opposite policies, the game is fully competitive.
Mixed games are stochastic games that are neither fully cooperative nor fully competitive.

\subsection{Policy Learning}

\iffalse
\begin{table}[!t]
\caption{Categorization of major RL algorithms.}
\begin{center}
\begin{tabular}{l|ll}
\Xhline{1.2pt}
 &  model-based & model-free \\
\Xhline{1.2pt}
value function & Dyna~\cite{sutton1991dyna} & \textit{Q}-learning~\cite{watkins1989learning} \\
 & Prioritized Sweeping~\cite{moore1993prioritized} & SARSA~\cite{rummery1994line}\\
 & \textit{Q}-iteration~\cite{busoniu2010reinforcement} & LSPI~\cite{lagoudakis2003least} \\
 & & DQN~\cite{mnih2015human}\\
\hline
policy search & REINFORCE~\cite{williams1992simple} & PILCO~\cite{deisenroth2011pilco} \\
 & GPOMDP~\cite{baxter2001infinite} & \\
 & Natural Policy Gradient~\cite{kakade2002natural} & \\
\hline
Hybrid & & Actor-Critic~\cite{konda2000actor} \\
 & & PGwFA~\cite{sutton2000policy} \\
 & & Natural Actor-Critic~\cite{peters2005natural}\\
 & & DPG~\cite{silver2014deterministic} \\
 & & DDPG~\cite{lillicrap2015continuous}\\
\Xhline{1.2pt}
\end{tabular}
\end{center}
\label{tab:RL_cate}
\end{table}
\fi

Reinforcement Learning is a class of learning problems in which the goal of an agent~(or multi-agent) to find the \textit{policy} to optimize some measures of its long-term performance.
RL solutions can be categorized in different ways. Here we investigate them from two perspectives: whether the full model is available and the way of finding the optimal policy.

\subsubsection{Model-based v.s. Model-free}
Reinforcement learning algorithms, which explicitly learn system models and use them to solve MDP problems, are model-based methods.
Model-based RL has a strong influence from the control theory and is often explained in terms of different disciplines.
These methods include popular algorithms such as the Dyna~\cite{DBLP:journals/sigart/Sutton91}, Prioritized Sweeping~\cite{DBLP:journals/ml/MooreA93}, \textit{Q}-iteration~\cite{busoniu2010reinforcement}, Policy Gradient~(PG)~\cite{DBLP:journals/ml/Williams92}, and the variation of PG~\cite{DBLP:journals/jair/BaxterB01,DBLP:conf/nips/Kakade01}.
The model-free methods ignore the model and just focus on figuring out the value functions directly from the interaction with the environment.
To accomplish this, the methods depend on sampling and observation heavily; thus they don't need to know the inner working of the system. Some examples of these methods are \textit{Q}-learning~\cite{DBLP:journals/ras/Krose95}, SARSA~\cite{rummery1994line}, LSPI~\cite{DBLP:journals/jmlr/LagoudakisP03}, and Actor-Critic~\cite{DBLP:conf/nips/KondaT99}.

\subsubsection{Value function v.s. Policy search}
The algorithms, which first find the optimal value functions and then extract optimal policies, are value function methods, such as Dyna, \textit{Q}-learning, SARSA, and DQN~\cite{DBLP:journals/nature/MnihKSRVBGRFOPB15}.
The alternative approaches are policy search methods which solve an MDP problem by directly searching in the space of policies.
An important class of policy search methods is that of Policy Gradient~(PG) algorithms~\cite{DBLP:journals/ml/Williams92,DBLP:journals/jair/BaxterB01,DBLP:conf/nips/Kakade01,DBLP:conf/icml/DeisenrothR11}.
These methods target at modeling and optimizing the policy directly. The policy is usually modeled with a parameterized function with respect to $\pi_{\theta}(a|s)$. The value of the reward (objective) function depends on this policy and then various algorithms can be applied to optimize $\theta$ for the best reward.
There are a series of algorithms, which use the PG to search in the policy space, and at the same time estimate a value function.
The important class of these methods are Actor-Critic~(AC) and its variation~\cite{DBLP:conf/nips/KondaT99,DBLP:conf/ecml/PetersVS05,DBLP:journals/ijon/PetersS08,DBLP:conf/nips/BhatnagarSGL07,DBLP:journals/automatica/BhatnagarSGL09}.
These are two-time-scale algorithms where the critic uses Temporal-Difference~(TD) learning with a linear approximation architecture and the actor is updated in an approximate gradient direction based on information provided by the critic.

\section{Reinforcement Learning for Search}
Search aims to find and rank a set of objects (e.g., documents, records) based on a user query~\cite{yin2016ranking}.
In this section, we review RL applications in key topics of search.

\subsection{Query Understanding}
Query understanding is the primary task for the search engine to understand users' information needs.
It can be potentially useful for improving general search relevance, user experience, and helping users to accomplish tasks~\cite{DBLP:journals/sigir/CroftBLX10}.
In~\cite{DBLP:conf/emnlp/NogueiraC17}, RL has been leveraged to solve the query reformulation task: a query reformulation framework is proposed based on a neural network, which rewrites a query to maximize the number of relevant documents returned.
In the proposed framework, a search engine is treated as a black box that an agent learns to use in order to retrieve more relevant items, which opens the possibility of training an agent to use a search engine for a task other than the one it was originally intended for.
Additionally, the upper-bound performance of an RL-based model is estimated in a given environment.
In~\cite{DBLP:journals/corr/abs-1809-10658}, a multi-agent based method is introduced to efficiently learn diverse query reformulation.
It is argued that it is easier to train multiple sub-agents than a single generalist one since each sub-agent only needs to learn a policy that performs well for a subset of examples.
In the proposed framework, an agent consists of multiple specialized sub-agents and a meta-agent that learns to aggregate the answers from sub-agents to produce a final answer.
Thus, the method makes learning faster with parallelism.

\subsection{Ranking}

Relevance Ranking is the core problem of information retrieval~\cite{DBLP:conf/kdd/YinHTDZOCKDNLC16} and learning to rank~(LTR) is the key technology in relevance ranking.
In LTR, the approaches to directly optimize the ranking evaluation measures are representative and have been proved to be effective\cite{DBLP:conf/sigir/YueFRJ07,DBLP:conf/sigir/XuL07,DBLP:conf/sigir/XuLLLM08}.
These methods usually only optimize the evaluation measure calculated at a predefined ranking position, e.g. NDCG at rank $K$ in~\cite{DBLP:conf/sigir/XuL07}. The information carried by the documents after the rank $K$ are neglected.
To solve such problem, in~\cite{DBLP:conf/sigir/WeiXLGC17}, an LTR model, MDPRank, is proposed based on Markov decision process, which has the ability to leverage the measures calculated at all of the ranking positions.
The reward function is defined based upon the IR evaluation measures and the model parameters can be learned through maximizing the accumulated rewards to all of the decisions.
Implicit relevance feedback refers to an interactive process between search engine and user, and has been proven to be very effective for improving retrieval accuracy~\cite{DBLP:conf/cikm/LvZ09}. Both Bandits and MDPs can model such an interactive process naturally~\cite{DBLP:conf/www/VorobevLGS15,DBLP:conf/icml/KatariyaKSW16,DBLP:conf/ijcai/KatariyaKSVW17}. In~\cite{DBLP:conf/icml/KvetonSWA15}, cascading bandits are introduced to identify the most attractive items, and the goal of the agent is to maximize its total reward with respect to the list of the most attractive items. Through maintaining state transition, MDP is able to model the user state in the interaction with search engine. 
In~\cite{DBLP:conf/ictir/ZengXLGC18}, the interactive process is formulated as an MDP and the Recurrent Neural Network is applied to process the feedback.

Beyond relevance ranking, another important goal is to provide search results that cover a wide range of topics for a query, i.e., search result diversification~\cite{DBLP:journals/ftir/SantosMO15,DBLP:journals/tist/XuXLGC17}.
Typical methods formulate the problem of constructing a diverse ranking as a process of greedy sequential document selection.
To select an optimal document for a position, it is critical for a diverse ranking model to capture the utility of information users have perceived from the preceding documents.
To explicitly model the utility perceived by the users, the construction of a diverse ranking is formalized as a process of sequential decision making and the process is modeled as a continuous state Markov decision process, referred to as MDP-DIV ~\cite{DBLP:conf/sigir/XiaXLGZC17}.
The ranking of $M$ documents is formalized as a sequence of $M$ decisions and each action corresponds to selecting one document from the candidate set.
In the parameter training phase, the policy gradient algorithm of REINFORCE is adopted and the expected long-term discounted rewards in terms of the diversity evaluation measure is maximized.
More works for diversity ranking see~\cite{DBLP:conf/sigir/FengXLGZC18,DBLP:conf/ijcai/KapoorKVC18}

\subsection{Whole-Page Optimization}

To improve user experiences, modern search engines aggregate versatile results from different verticals -- web-pages, news, images, video, shopping, knowledge cards, local maps, etc.
Page presentation is broadly defined as the strategy to present a set of items on search result page~(SERP), which is much more expressive than a ranked list.
Finding proper presentation for a gallery of heterogeneous results is critical for modern search engines.
One approach of efficiently learning to optimize a large decision space is fractional factorial design.
However, the method could cause a combinatorial explosion problem with a large search space.
In~\cite{DBLP:conf/kdd/HillNLIV17},  bandit formulation is applied to explore the layout space efficiently and hill-climbing is used to select optimal content in real-time.
The model avoids a combinatorial explosion in model complexity by only considering pairwise interactions between page components.
This approach is a greedy alternating optimization strategy that can run online in real-time.
In~\cite{DBLP:conf/wsdm/WangYJWYCM16,DBLP:journals/tweb/WangYJWYCM18}, a framework is proposed to learn the optimal page presentation to render heterogeneous results onto SERP.
It leveraged the MDP setting and the agent is designed as the algorithm that determines the presentation of page content on a SERP for each incoming search query.
To solve the critical efficiency problem, it proposed a policy-based learning method which can rapidly choose actions from the high-dimensional space.

\subsection{Session Search}

The task-oriented search includes a series of search iterations triggered by the query reformulations within a session.
Markov chain in session search is observed: user’s judgment of search results in the prior iteration will influence user’s behaviors in the next search iteration.
Session search is modeled as a dual-agent stochastic game based on Partially Observable Markov Decision Process (POMDP) in~\cite{DBLP:conf/sigir/LuoZY14}.
They mathematically model dynamics in session search as a cooperative game between the user and the search engine, while user and the search engine work together in order to jointly maximize the long-term cumulative rewards.
Log-based document re-ranking is a special type of session search that re-ranks documents based on the historical search logs which includes the target user's personalized query log and other users' search activities.
The re-ranking aims to offer a better order of the initial retrieved documents~\cite{DBLP:conf/sigir/ZhangLY14}.
Nowadays, deep reinforcement learning technology has been applied in the E-Commerce search engine~\cite{DBLP:conf/kdd/HuDZ0X18,DBLP:conf/www/FengLHLOWZ18}.
For better utilizing the correlation between different ranking steps, RL is used to learn an optimal ranking policy which maximizes the expected accumulative rewards in a search session~\cite{DBLP:conf/kdd/HuDZ0X18}.
It formally defined the multi-step ranking problem in the search session as MDP, denoted as SSMDP, and proposed a novel policy gradient algorithm for learning an optimal ranking policy, which is able to deal with the problem of high reward variance and unbalanced reward distribution.
In~\cite{DBLP:conf/www/FengLHLOWZ18}, multi-scenario ranking is formulated as a fully cooperative, partially observable, multi-agent sequential decision problem, denoted as MA-RDPG.
MA-RDPG has a communication component for passing message, several private agents for making action for ranking, and a centralized critic for evaluating the overall performance of the co-working agents.
Agents collaborate with each other by sharing a global action-value function and passing messages that encode historical information across scenarios.
\section{Reinforcement Learning for Recommendation}
Recommender systems target to capture users' preferences according to their feedback~(or behaviors, e.g. rating and review) and suggest items that match their preferences. In this section, we briefly review how RL is adapted in several key tasks in recommendations. 

\subsection{Exploitation/Exploration Dilemma}
Traditional recommender systems suffer from the exploitation-exploration dilemma, where exploitation is to recommend items that are predicted to best match users' preferences, while exploration is to recommend items randomly to collect more users' feedback. The contextual bandit models an agent that attempts to balance the competing exploitation and exploration tasks in order to maximize the accumulated long-term reward over a considered period. The traditional strategies to balance exploitation and exploration in bandit setting are $\epsilon$-greedy~\cite{watkins1989learning}, EXP3~\cite{DBLP:journals/siamcomp/AuerCFS02}, and UCB1~\cite{DBLP:journals/ml/AuerCF02}. In the news feeds scenario, the exploration/exploitation problem of personalized news recommendation is modeled as a contextual bandit problem~\cite{DBLP:conf/www/LiCLS10}, and a learning algorithm LinUCB is proposed to select articles sequentially for specific users based on the users' and articles' contextual information, in order to maximize the total user clicks. 

\subsection{Temporal Dynamics}

Most existing recommender systems such as collaborative filtering, content-based and learning-to-rank have been extensively studied with the stationary environment (reward) assumption, where user's preference is assumed to be static. However, this assumption is usually not true in reality since users’ preferences are dynamic, thus the reward distributions usually change over time. In bandit setting, it usually introduces a variable reward function to delineate the dynamic nature of the environment. For instance, the particle learning based dynamical context drift model is proposed to model the changing of reward mapping function in multi-armed bandit problem, where the drift of the reward mapping function is learned as a group of random walk particles, and well fitted particles are dynamically chosen to describe the mapping function~\cite{DBLP:conf/kdd/ZengWML16}.  A contextual bandit algorithm is presented to detect the changes of environment according to the reward estimation confidence, and updates the arm selection policy accordingly~\cite{DBLP:conf/sigir/WuIW18}. The change-detection based framework under the piecewise-stationary reward assumption for the multi-armed bandit problem is proposed in~ \cite{DBLP:conf/aaai/LiuLS18}, where upper confidence bound (UCB) policies is used to detect change points actively and restart the UCB indices.  Another solution for capturing user's dynamic preference is to introduce the MDP setting~\cite{DBLP:conf/kdd/Chen0DTHT18,DBLP:journals/corr/abs-1810-12027,zhao2018recommendations,DBLP:conf/dasfaa/zoulixin}. Under the MDP setting, \textit{state} is introduced to represent user's preference and \textit{state transition} captures the dynamic nature of user's preference over time. In~\cite{zhao2018recommendations},  a user's dynamic preference (agent's state) is learned from his/her browsing history. Each time the recommender system suggests an item to a user, the user will browse this item and provide feedback (skip, click or purchase), which reveals user's satisfaction of the recommended item. According to the feedback, the recommender system will update its state to represent user's new preferences~\cite{zhao2018recommendations}. 

%To solve the unstable reward distribution problem in dynamic recommendation environments, approximate regretted reward technique is proposed with Double DQN to obtain a reference baseline from individual  customer sample, which can effectively stabilize the reward value estimation and enhance the recommendation quality

%A Change-Detection Based Framework for Piecewise-Stationary Multi-Armed Bandit Problem \cite{DBLP:conf/aaai/LiuLS18}.
%Learning Contextual Bandits in a Non-stationary Environment \cite{DBLP:conf/sigir/WuIW18}.
%Iterative Model Refinement of Recommender MDPs based on Expert Feedback \cite{DBLP:conf/pkdd/KhanPA13}
%Stabilizing Reinforcement Learning in Dynamic Environment with Application to Online Recommendation \cite{DBLP:conf/kdd/Chen0DTHT18}.
%Recommendations with Negative Feedback via Pairwise Deep Reinforcement Learning \cite{DBLP:conf/kdd/ZhaoZDXTY18}.
%Deep Reinforcement Learning based Recommendation with Explicit User-Item Interactions Modeling \cite{liu2018deep}

\subsection{Long Term User Engagement}
User engagement in recommendation is the assessment of user's desirable (even essential) responses to the items (products, services, or information) suggested by the recommender systems \cite{DBLP:series/synthesis/2014Lalmas}. User engagement can be measured not only in terms of immediate response (e.g. clicks and rating of the recommended items), but more importantly in terms of long-term response (e.g. user repetitively purchases)~\cite{DBLP:conf/recsys/SchopferK14}. In~\cite{DBLP:conf/cikm/WuWHS17}, the problem of long-term user engagement optimization is formulated as a sequential decision making problem. In each iteration, the agent needs to estimate the risk of losing a user based on the user's dynamic response to past recommendations. Then, a bandit based method~\cite{DBLP:conf/cikm/WuWHS17} is introduced to balance the immediate user click and the expected future clicks when the user revisits the recommender system. In practical recommendation sessions, users will sequentially access multiple scenarios, such as the entrance pages and the item detail pages, and each scenario has its own recommendation strategy. A multi-agent reinforcement learning based approach (DeepChain) is proposed in~\cite{zhao2019model}, which can capture the sequential correlation among different scenarios and jointly optimize multiple recommendation strategies. To be specific, model-based reinforcement learning technique is introduced to reduce the training data requirement and execute more accurate strategy updates. In the news feeds scenario~\cite{DBLP:conf/www/ZhengZZXY0L18}, to incorporate more user feedback information, the long-term user response (i.e., how frequent user returns) is considered as a supplement to user's immediate click behaviors, and a Deep Q-Learning based framework is proposed to optimize the news recommendation strategies. 
%A Unified Contextual Bandit Framework for Longand Short-Term Recommendations \cite{DBLP:conf/pkdd/TavakolB17},
%DRN: A Deep Reinforcement Learning Framework for News Recommendation \cite{DBLP:conf/www/ZhengZZXY0L18}
%Returning is Believing: Optimizing Long-term User Engagement in Recommender Systems\cite{DBLP:conf/cikm/WuWHS17}
%Long Term Recommender Benchmarking for Mobile Shopping List Applications using Markov Chains \cite{DBLP:conf/recsys/SchopferK14}

\subsection{Page-Wise Recommendation}
In practical recommender systems, each time users are typically recommended a page of items. In this setting, the recommender systems need to jointly (1) select a set of complementary and diverse items from a larger candidate item set and (2) form an item display (layout configuration) strategy to place the items in a 2-D web page that can lead to maximal reward. Given the massive number of items, the action space is extremely large if we treat each whole page recommendation as one action. To mitigate the issue of the large action space, a Deep Deterministic Policy Gradient algorithm is proposed~\cite{dulac2015deep} where the Actor generates a deterministic optimal action according to the current state, and the Critic outputs the Q-value of this state-action pair. DDPG reduces the computational cost of conventional value-based reinforcement learning methods, thus it is a fitting choice for the whole page recommendation setting~\cite{DBLP:conf/www/CaiFTZ18,DBLP:conf/aaai/CaiFTZ18}. Several approaches are presented recently to enhance the efficiency~\cite{DBLP:journals/corr/abs-1801-05532,DBLP:journals/corr/abs-1811-05869}. In~\cite{zhao2018deep,zhao2017deep}, CNN techniques are introduced to capture the item display patterns and users' feedback of each item in the page. To represent each item, item-embedding, category-embedding and feedback embedding are leveraged, which can help to generate complementary and diverse recommendations and capture user's interests within the pages. Bandit techniques are also leveraged for whole-page Recommendations~\cite{DBLP:conf/aaai/WangOWCAC17,DBLP:journals/ijon/Lacerda17}. For instance, the whole page recommendation task is considered as a combinatorial semi-bandit problem, where the system recommends $S$ actions from a candidate set of $K$ actions, and displays the selected items in $S$ (out of $M$) positions~\cite{DBLP:conf/aaai/WangOWCAC17}. 

%A Thompson sampling algorithm is derived for the contextual combinatorial problem with a minimum-cost maximum-flow network, which solves the computational intractability problem of bandit setting.} {\bf need to give at least one example???}

%Deep reinforcement learning for page-wise recommendations \cite{DBLP:conf/recsys/ZhaoXZDYT18}
%Deep Reinforcement Learning for List-wise Recommendations \cite{DBLP:journals/corr/abs-1801-00209}
%Efficient Ordered Combinatorial Semi-Bandits for Whole-Page Recommendation \cite{DBLP:conf/aaai/WangOWCAC17}.
%Multi-Objective Ranked Bandits for Recommender Systems \cite{DBLP:journals/ijon/Lacerda17}
%Reinforcement Mechanism Design for e-commerce \cite{DBLP:conf/www/CaiFTZ18}
%Reinforcement Mechanism Design for Fraudulent Behaviour in e-Commerce \cite{DBLP:conf/aaai/CaiFTZ18}.
%Reinforcement Learning based Recommender System using Biclustering Technique \cite{choi2018reinforcement}
%Large-scale Interactive Recommendation with Tree-structured Policy Gradient \cite{chen2018large}
\section{Reinforcement Learning for Online Advertising}
The goal of online advertising is to assign the right advertisements to the right users so as to maximize the revenue, click-through rate (CTR) or return on investment~(ROI) of the advertising campaign.
The two main marketing strategy in online advertising are guaranteed delivery~(GD) and real-time bidding~(RTB).

\subsection{Guaranteed Delivery}
In guaranteed delivery, advertisements that share a single idea and theme are grouped into campaigns, and are charged on a pay-per-campaign basis for the pre-specified number of deliveries~(click or impressions)~\cite{DBLP:conf/cikm/SalomatinLY12}.
Most popular GD~(Guaranteed Delivery) solutions are based on offline optimization algorithms, and then adjusted for online setup.
However, deriving the optimal strategy to allocate impressions is challenging, especially when the environment is unstable in real-world application.
In~\cite{DBLP:journals/corr/abs-1809-03152}, a multi-agent reinforcement learning~(MARL) approach is proposed to derive cooperative policies for the publisher to maximize its target in an unstable environment.
They formulated the impression allocation problem as an auction problem where each contract can submit virtual bids for individual impressions.
With this formulation, they derived the optimal impression allocation strategy by solving the optimal bidding functions for contracts.

\subsection{Real-Time Bidding}
RTB allows an advertiser to submit a bid for each individual impression in a very short time frame. Ad selection task is typically modeled as multi-armed bandit (MAB) problem with the setting that samples from each arm are iid, feedback is immediate and rewards are stationary~\cite{DBLP:conf/icdm/YangL16,DBLP:conf/aaai/NuaraT0R18,DBLP:conf/ijcnn/GaspariniNT0R18,DBLP:conf/cikm/TangRSA13,DBLP:conf/nips/XuQL13,DBLP:journals/corr/YuanWM13,DBLP:journals/mktsci/SchwartzBF17}. The payoff functions of a MAB are allowed to evolve, but they are assumed to evolve slowly over time. On the other hand, display ads created while others are removed regularly in an advertising campaign circulation.
The problem of multi-armed bandits with budget constraints and variable costs is studied in~\cite{DBLP:conf/aaai/DingQZL13}.
In this case, pulling the arms of bandit will get random rewards with random costs, and the algorithm aims to maximize the long-term reward by pulling arms with a constrained budget.
This setting can model Internet advertising in a more precise way than previous works where pulling an arm is costless or has a fixed cost.

Under the  MAB setting,  the bid decision is considered as a static optimization problem of either treating the value of each impression independently or setting a bid price to each segment of ad volume. However, the bidding for a given ad campaign would repeatedly happen during its life span before the budget running out. Thus, the MDP setting have also been studied~\cite{DBLP:conf/wsdm/CaiRZMWYG17,DBLP:conf/ijcai/Tang17a,DBLP:journals/corr/abs-1809-03149,DBLP:conf/kdd/0009QG0H18,DBLP:journals/corr/abs-1808-00720,DBLP:conf/cikm/WuCYWTZXG18,DBLP:conf/cikm/JinSLGWZ18}.
A model-based reinforcement learning framework is proposed to learn bid strategies in RTB advertising~\cite{DBLP:conf/wsdm/CaiRZMWYG17}, where neural network is used to approximate the state value, which can better deal with the scalability problem of large auction volume and limited campaign budget.
A model-free deep reinforcement learning method is proposed to solve the bidding problem with constrained budget~\cite{DBLP:conf/cikm/WuCYWTZXG18}: the problem is modeled as a $\lambda$-control problem, and RewardNet is designed for generating rewards to solve reward design trap, instead of using the immediate reward.
A multi-agent bidding model is presented, which takes the other advertisers' bidding in the system into consideration, and a clustering approach is introduced to solve the large number of advertisers challenge~\cite{DBLP:conf/cikm/JinSLGWZ18}.

\iffalse
\subsection{CTR prediction}
In online advertising, click-through rate (CTR) is a very important metric for evaluating ad performance.
As a result, click prediction systems are essential and widely used for guaranteed delivery and real-time bidding.

RecoGym is Reinforcement Learning environment for online advertising to model the intrinsic dimensions of user/item clusters and vary the influence of repeated exposure on ad CTR~\cite{DBLP:journals/corr/abs-1808-00720}

In online display advertising, visual design is important: changing of layout of ads can influence it effectiveness, e.g., the total revenue or click-through rate~\cite{DBLP:conf/cikm/TangRSA13}. Deciding which layout to use involves a trade-off between exploitation and exploration, ,i.e., leveraging an effective layout we already know, or trying a new layout. In order to balance exploitation and exploration, a contextual bandit is designed for the automatic layout selection problem, and offline replay is introduced as an estimator to evaluate the performance of ad layouts. There are several other bandits based models proposed for performance evaluation task. For instance, , and bias-correction methods are proposed to correct the “estimation of the largest mean” (ELM) bias problem which results from search engine always selecting the most profitable ads to display~\cite{DBLP:conf/nips/XuQL13}. Other applications of bandit based techniques for Internet advertising includes keywords extraction~\cite{DBLP:journals/corr/YuanWM13} and target user discovery~\cite{DBLP:journals/mktsci/SchwartzBF17}.
\fi
\section{Conclusion and Future Directions}
In this article, we present an overview of information seeking from the reinforcement learning perspective. We first introduce mathematical foundations of RL based information seeking approaches. Then we review state-of-the-art algorithms of three representative information seeking mechanisms -- search, recommendations, and advertising. Next, we here discuss some interesting research directions on reinforcement learning that can bring the information seeking research into a new frontier.

First, most of the existing works train a policy within one scenario, while overlooking users' behaviors (preference) in other scenarios~\cite{DBLP:conf/www/FengLHLOWZ18}. This will result in a suboptimal policy, which calls for collaborative RL frameworks that consider search, recommendation and advertising scenarios simultaneously. Second, the type of reward function varies among different computational tasks. More sophisticated reward functions should be designed to achieve more goals of information seeking, such as increasing the supervising degree of recommendations. Third, more types of user-agent interactions could be incorporated into RL frameworks, such as adding items into shopping cart, users' repeat purchase behavior, users' dwelling time in the system, and user's chatting with customer service representatives or agent of AI dialog system. Fourth, testing a new algorithm is expensive since it needs lots of engineering efforts to deploy the algorithm in the practical system, and it also may have negative impacts on user experience if the algorithm is not mature. Thus online environment simulator or offline evaluation method based on historical logs are necessary to pre-train and evaluate new algorithms before launching them online. Finally, there is an increasing demand for an open online reinforcement learning environment for information seeking, which can advance the RL and information seeking communities and achieve better consistency between offline and online performance.

\section*{ACKNOWLEDGEMENTS}
Xiangyu Zhao and Jiliang Tang are supported by the National Science Foundation (NSF) under grant numbers IIS-1714741, IIS-1715940 and CNS-1815636, and a grant from Criteo Faculty Research Award. 

\bibliographystyle{sigwebnewsletter} 
\nocite{*}
\bibliography{instructions}
\begin{biography}
Xiangyu Zhao is a Ph.D. student of computer science and engineering at Michigan State University (MSU). His supervisor is Dr. Jiliang Tang. Before joining MSU, he completed his MS(2017) at USTC and BS(2014) at UESTC. He is the student member of IEEE, SIGIR, and SIAM. His current research interests include data mining and machine learning, especially (1) Reinforcement Learning for E-commerce; (2) Urban Computing and Spatio-Temporal Data Analysis. After joining MSU, he has published his work in top journals (e.g. SIGKDD Explorations) and conferences (e.g., KDD, ICDM, CIKM, RecSys). He was the recipients of the RecSys’18, KDD’18, SDM’18, and CIKM’17 Student Travel Award.

Long Xia is a research scientist in Data Science Lab at JD.com. He is now mainly responsible for applying advanced technology to the E-commerce recommender system in JD.com. Before that, he received his Ph.D. in Computer Science from Institute of Computing Technology, Chinese Academy of Sciences. His research interests include data mining, applied machine learning, information retrieval, and recommender system. He has published his research in top journals and conferences, e.g. TIST, SIGIR, KDD, RecSys.

Jiliang Tang is an assistant professor in the computer science and engineering department at Michigan State University. Before that, he was a research scientist in Yahoo Research and got his PhD from Arizona State University in 2015. He has broad interests in social computing, data mining and machine learning. He was the recipients of the Best Paper Award in ASONAM 2018, the Best Student Paper Award in WSDM2018, the Best Paper Award in KDD2016, the runner up of the Best KDD Dissertation Award in 2015, Dean's Dissertation Award and the best paper shortlist of WSDM2013. He is now associate editors of ACM TKDD, ICWSM and Neurocomputing. He has published his research in highly ranked journals and top conference proceedings, which received thousands of citations and extensive media coverage. 

Dawei Yin is Senior Director at JD.com, leading the science efforts of recommendation, search, metrics and knowledge graph. Prior to joining JD.com, he was Senior Research Manager at Yahoo Labs, leading relevance science team and in charge of Core Search Relevance of Yahoo Search. He obtained Ph.D. (2013), M.S. (2010) from Lehigh University and B.S. (2006) from Shandong University. His research interests include data mining, applied machine learning, information retrieval and recommender system. He published tens of research papers in premium conferences and journals, and won WSDM2016 best paper award, KDD2016 best paper award, and WSDM2018 best student paper award.

\end{biography}
\end{document}